\documentclass[prc,twocolumn,superscriptaddress,showpacs,floatfix]{revtex4}

\usepackage{graphicx}
\usepackage{dcolumn}
\usepackage{bm}
\begin{document}

\title{Graphical Method for Effective Interaction with a New Vertex Function}

\author{K. Suzuki}
\affiliation{
Senior Academy,
Kyushu Institute of Technology,
Kitakyushu 804-8550, Japan
}

\author{R. Okamoto}
\email{okamoto@mns.kyutech.ac.jp}
\affiliation{
Department of Physics,
Kyushu Institute of Technology,
Kitakyushu 804-8550, Japan
}

\author{H. Kumagai}
\email{kumagai@fit.ac.jp}
\affiliation{
Faculty of Information Engineering,
Fukuoka Institute of Technology,
Fukuoka 811-0295, Japan
}

\author{S. Fujii}
\email{sfujii@cns.s.u-tokyo.ac.jp}
\affiliation{
Center for Nuclear Study (CNS),
University of Tokyo,
Wako Campus of RIKEN,
Wako 351-0198, Japan
}


\date{\today}
\begin{abstract}
Introducing a new vertex function, $\hat{Z}(E)$, of an energy variable $E$, we derive a new equation for the effective interaction.
The equation is obtained by replacing the $\hat{Q}$-box in the Krenciglowa-Kuo (KK) method by  $\hat{Z}(E)$. 
This new approach
can be viewed as an extension of the KK method. We show that this equation can be solved both in iterative and non-iterative ways.
We observe that the iteration procedure with $\hat{Z}(E)$ brings about fast convergence compared to the usual KK method. It is shown that, as in the KK approach, the procedure of calculating the effective interaction can be reduced to determining the true eigenvalues of the 
original Hamiltonian $H$ and they can be obtained as the positions of intersections of graphs generated from $\hat{Z}(E)$.
We find that this graphical method yields always precise results and reproduces any of the true eigenvalues of $H$.
The calculation in the present approach can be made regardless of  overlaps with the model space
 and energy differences 
between unperturbed energies and the eigenvalues of $H$. We find also that $\hat{Z}(E)$ is a well-behaved function of $E$ and has no singularity. These characteristics of the present approach ensure stability in actual calculations and would be helpful to resolve some difficulties due to the presence of poles in the $\hat{Q}$-box. Performing  test calculations, we verify numerically  theoretical predictions made in the present approach.

\end{abstract}

\maketitle
\section{\label{sec:introduction}Introduction}
In nuclear, atomic, and chemical physics, it is often useful to introduce an effective interaction acting in a chosen model space (P space) of  tractable dimension.
In nuclear physics, much effort has been made both as regards formal theories and their
applications~\cite{Kuo-Osnes90,Ellis-Osnes77,Navratil-a,Navratil-b,Hjorth-Jensen,SOK87,Andreozzi,OFS05}. Recently the effective interaction method has been applied to new fields of many-body physics, such as 
quantum dots~\cite{Varga,quantumdot} and many-boson systems~\cite{bosons}.
 
Among many approaches, we here direct our attention to the Krenciglowa-Kuo (KK)~\cite{KK74,KKSO95} and the Lee-Suzuki (LS)~\cite{Lee-Suzuki80,Suzuki-Lee80} methods.
These two methods are constructed in terms of the so called $\hat{Q}$-box as a building block of formulation.
The KK approach has very simple structure and the effective interaction is obtained in an iterative way.
If the iteration converges, in almost all numerical calculations, eigenvalues are given for the states which have the largest overlaps with the chosen model space.
On the other hand, the LS method reproduces eigenvalues for the states which lie closest to the chosen unperturbed energy. Originally the LS method had been presented to resolve the difficulty of divergence in the perturbation expansion.
The LS method is rather complicated in structure and higher derivatives of the $\hat{Q}$-box with respect to starting energy are necessary if one wishes to obtain more accurate solutions.

Both of the two theories yield only certain of the true eigenvalues of the original Hamiltonian.
This restriction is not desirable.
In a formal point of view, the $\hat{Q}$-box itself contains information regarding all of the true eigenvalues.
For a given model space of dimension $d$, there would be a method of reproducing any $d$ eigenvalues among all the true eigenvalues.

In many cases of applying the effective interaction theories an iteration  or a recursion method has often been  employed.
In general, the convergence in such a method depends strongly on the properties of the eigenstates of 
the Hamiltonian $H$.
In actual calculations the information on the true eigenstates is not given beforehand. Therefore, it is impossible to 
control the convergence in the iteration. In many cases we cannot know whether the iteration is convergent or not before starting calculations. Even if the iteration converges, we do not know which eigenvalues are reproduced in all of the eigenvalues.
Another difficulty encountered in actual calculations is the pole problem. The $\hat{Q}$-box itself has poles at the energies which are the
eigenvalues of $H$ in the complement space (Q space).
 The presence of poles causes often instability in numerical calculations.

We shall show  that it is indeed possible to resolve these difficulties by introducing  a new vertex function, $\hat{Z}(E)$, 
in place of the $\hat{Q}$-box by Kuo {\it et al.}~\cite{KK74,KKSO95}.
Preliminary version of the present work has been reported in Ref.~\cite{OSKF2010}. 
Very recently Dong, Kuo, and Holt have 
followed the present approach and applied to the actual calculations of the shell-model effective 
interactions~\cite{DKH2010}.
These works have shown that the present method has a possibility of providing a suitable framework 
with $\hat{Z}(E)$ as well as the KK approach.

The organization of the present paper is as follows:
In Sec.~\ref{formulation} we outline the $\hat{Q}$-box formalism and the KK method. 
A new vertex function operator, $\hat{Z}(E)$, is introduced and some of its mathematical properties are clarified. The algorithm of the
calculation procedure is given by applying the  secant and the Newton-Raphson methods. In Sec.~\ref{test} test calculations are made
 in order to assess the present approach. We examine whether the theoretical predictions are verified or not 
 in an exactly solvable model. Concluding remarks are given in Sec.~\ref{concluding}.
\section{\label{formulation}Formulation}
We consider  a general quantum system which is described by a Hamiltonian $H$.
We write an eigenvalue equation with the  eigenvalues $\{ E_{k}\}$ and the eigenstates $\{ |\Psi_{k} \rangle\}$ as
\begin{eqnarray}
\label{eq:true-eigen-eq}
 H |\Psi_{k}\rangle=E_{k} |\Psi_{k}\rangle, 
 \quad \quad k = 1, 2, \cdots. 
\end{eqnarray}
The Hamiltonian $H$ is  supposed to be composed of the unperturbed Hamiltonian $H_0$ and the perturbation $V$, {\it i.e.}, $H=H_0+V$. 
We decompose the entire Hilbert space into the model space (P space)  and its complement (Q space) with the projection operators $P$ and $Q$, respectively.
Basic properties of the projection operators are $P+Q=1$ and $PQ=QP=0$.

We here assume that $H_0$ is decoupled between the P and Q spaces as 
\begin{equation}
\label{eq:Hamiltonian}
H_0=PH_0P+QH_0Q.
\end{equation}
In the present work we consider a case that the P-space states have a degenerate unperturbed energy $E_0$, {\it i.e.},
\begin{equation}
\label{eq:model-space-hamiltonian}
PH_0P=E_0P.
\end{equation}
Then the P-space eigenvalue equation is written with the effective interaction $R$ 
and the eigenstate $|\phi_{k}\rangle$ as 
\begin{eqnarray}
\label{eq:p-space-eigen}
 (E_0 P + R)|{\phi_k }\rangle =E_k | {\phi _k } \rangle, 
 \quad \quad k = 1, 2, \cdots, d,
\end{eqnarray}
where $d$ is the dimension of the P space.
In the above $d$ eigenvalues $E_k$'s should agree with $d$ of the true eigenvalues in 
Eq.~(\ref{eq:true-eigen-eq}). Various solutions for $R$ are possible, and many theoretical frameworks have been given for obtaining $R$. Probably the most
widely applied effective interaction is given by imposing the condition that the model-space eigenstate
$|\phi_{k}\rangle$ in Eq.~(\ref{eq:p-space-eigen}) should agree with the P-space component of the true eigenstate $|\Psi_{k}\rangle$
of $H$, {\it i.e.}, $|\phi_{k}\rangle=P|\Psi_{k}\rangle$. This restriction on $|\phi_{k}\rangle$ leads to the standard non-Hermitian form of $R$~\cite{Brandow}.

\subsection{\label{solution}Solutions for effective interaction in the $\hat{Q}$-box formalism}
Among many approaches to the effective interaction $R$ we here discuss the KK formalism~\cite{KK74,KKSO95}. 
Originally the KK method is based on a diagrammatic representation of the effective interaction, which has been known as the $\hat{Q}$-box folded-diagram method originated by Kuo {\it et al.}~\cite{KLR71}.

For obtaining  $R$, one first calculates the vertex function called  the $\hat{Q}$-box which is defined as
 the sum of all the linked and non-folded diagrams.
Next one should add the folded diagrams, which can be carried out rather simply 
by applying the energy-derivative expression of the $\hat{Q}$-box~\cite{KK74,KLR71}.
 Originally the KK approach was proposed to derive an effective
interaction acting among a few valence particles outside the closed-shell core.
However, in the present work, we consider a general quantum system and wish to reproduce $d$ total energies
of $H$ by introducing an effective interaction. For this case the KK method can also
be applied by defining the $\hat{Q}$-box in an operator form as
\begin{eqnarray}
\label{eq:Q-box}
 \hat Q(E)\equiv PVP + PVQ\frac{1}{{E - QHQ}}QVP,
 \end{eqnarray}
which is a function of an energy variable $E$. Equation~(\ref{eq:Q-box}) is equivalent to the
energy-dependent form of the effective interaction given by Bloch and Horowitz
 in the many-body perturbation theory~\cite{Bloch}.

With the $\hat{Q}$-box the effective interaction $R$ can be expanded into
\begin{eqnarray}
\label{eq:R-expansion}
R=\hat{Q}+\hat{Q}_{1}\hat{Q}+\hat{Q}_{1}\hat{Q}_{1}\hat{Q}+\hat{Q}_{2}\hat{Q}\hat{Q}+\cdots,
\end{eqnarray}
where $\hat{Q}\equiv \hat{Q}(E_{0})$ and $\hat{Q}_{m}\equiv \hat{Q}_{m}(E_{0})$ with
\begin{eqnarray}
\label{eq:Q-E0-derivative}
\hat{Q}_{m}(E)\equiv\frac{1}{m!} 
                             \frac{d^{m}\hat{Q}(E)}{dE^{m}}, \ \ m=1, 2, \cdots.
\end{eqnarray}
Here $E_{0}$ is the starting energy or the degenerate unperturbed energy as given in 
Eq.~(\ref{eq:model-space-hamiltonian}).

If the series expansion in Eq.~(\ref{eq:R-expansion}) converges, $R$ is given in a formal  way by 
\begin{eqnarray}
\label{eq:RKK-1}
 R=\sum\limits_{k = 1}^d {\hat Q} (E_k )|{\phi_k}\rangle\langle {\tilde\phi_k}|,
\end{eqnarray}
where $E_{k}$ and $|\phi_{k}\rangle$ are given in Eq.~(\ref{eq:p-space-eigen}), and 
$\langle\tilde\phi_{k}|$ is the biorthogonal state defined through the orthogonality 
$\langle{\tilde\phi_k |\phi _{k'} }\rangle  = \delta_{kk'}$.
We here note that, as seen in Eqs.~(\ref{eq:p-space-eigen}) and (\ref{eq:RKK-1}),  the derivation of the effective interaction $R$ 
is equivalent to determining $d$ eigenvalues \{$E_k$\} and the corresponding P-space eigenstates $\{ |\phi_{k} \rangle\}$.
Using Eqs.~(\ref{eq:p-space-eigen}) and (\ref{eq:RKK-1}), the effective interaction $R$ is rewritten simply as 
\begin{eqnarray}
\label{eq:RKK-2}
R=\sum\limits_{k = 1}^d (E_{k}-E_{0})|{\phi_k}\rangle \langle {\tilde\phi_k}|,
 \end{eqnarray}
and the model-space eigenvalue equation with the 
effective interaction $R$ is expressed as
\begin{eqnarray}
\label{eq:p-space-eigen-eq-2}
[E_{0}P+{\hat Q} (E_k )]|{\phi_k}\rangle=E_{k}|{\phi_k}\rangle. 
 \end{eqnarray}
From Eqs.~(\ref{eq:RKK-1}) and (\ref{eq:p-space-eigen-eq-2}) we understand  that $R$ can be given by
calculating the $\hat{Q}$-box at the true eigenvalues $\{  E_k \}$ of $H$ as starting energies which are determined self-consistently.

We here show that there is another way of deriving the effective interaction $R$. 
We first note that the $\hat{Q}$-box is a function of  $E$ and resultantly the eigenvalues of $E_{0}P+\hat{Q}(E)$ are also  functions of $E$. We write the eigenvalue 
equations in the P space for an arbitrary energy $E$  as 
\begin{eqnarray}
\label{eq:Q-box-eigen-G}
[E_0 P+ \hat Q(E)] |\psi_{m}\rangle=G_{m}(E) | \psi_{m}\rangle, m = 1, 2, \cdots, d.
\end{eqnarray}
Since the P space is $d$ dimensional, we have $d$ eigenvalues denoted by $\{ G_{m}(E); m=1, 2, \cdots, d\}$, 
which we label in order of energy as $G_{1}(E) < G_{2}(E)<\cdots < G_{d}(E)$.
It may be clear from Eqs.~(\ref{eq:p-space-eigen-eq-2}) and (\ref{eq:Q-box-eigen-G})
 that the eigenvalues 
\{$E_k$\} in Eq.~(\ref{eq:p-space-eigen-eq-2}) can be given by solving the following equations 
\begin{equation}
\label{eq:EGK}
G_{m}(E)=E, \ \ m=1, 2, \cdots, d.
\end{equation}
Various mathematical methods have been known to solve such equations.

We should note that a set of the equations (\ref{eq:EGK})  are independent of the properties of the eigenstates of $H$, such as P-space overlaps or energy spacings. Therefore, in principle, it is possible to reproduce the true
eigenvalues $\{ E_k\}$ of $H$ more than in the usual KK method based on Eq.~(\ref{eq:p-space-eigen-eq-2}). However, as seen in  Eq.~(\ref{eq:Q-box}),
there appear poles in the $\hat{Q}$-box when $E$ approaches one of the eigenvalues of $QHQ$. The poles
in $\hat{Q}(E)$ induce also the poles in $G_m (E)$ in Eq.~(\ref{eq:Q-box-eigen-G}). 
Such a situation causes instability in numerically solving Eq.~(\ref{eq:EGK}) for $\{ E_k \}$
around the pole positions.
\subsection{\label{extention}Extension of the Krenciglowa-Kuo method}
In order to resolve the pole problem we introduce a new vertex function  of 
an  energy variable $E$, which is a P-space operator defined in terms of the $\hat{Q}$-box  and its energy derivative as~\cite{OSKF2010}
\begin{eqnarray}
\label{eq:z-operator}
 \hat Z(E)\equiv\frac{1}{{1 - \hat Q_1 (E)}}\left[ {\hat Q(E) -\hat Q_1 (E)(E - E_0 )P} \right].
 \end{eqnarray}
Hereafter we  shall refer  to $\hat Z(E)$ as the $\hat{Z}$-box.
We note here that the $\hat{Z}$-box agrees, at $E=E_0$, with the first-order
 recursive solution in the LS method~\cite{Suzuki-Lee80}.
In a recent paper of Dong, Kuo, and Holt~\cite{DKH2010} the definition of $\hat{Z}(E)$ is given  in a more general case with a non-degenerate P-space
unperturbed Hamiltonian $PH_{0}P$. In this case one should replace $E_0P$ by $PH_0P$.
As in Eq.~(\ref{eq:Q-box-eigen-G}) with the $\hat{Q}$-box, we consider an eigenvalue problem 
\begin{eqnarray}
\label{eq:Z-box-eigen}
 [E_0 P+ \hat{Z}(E)] |\psi_{m}\rangle=F_{m}(E) | \psi_{m}\rangle,\ \ m=1, 2, \cdots, d,
\end{eqnarray}
where $\{ F_{m}(E); m=1, 2, \cdots, d\}$ are  $d$ eigenvalues which are  functions of $E$. 
We here label $\{ F_{m}(E); m=1, 2, \cdots, d\}$ in order of  energy as
 $F_{1}(E) < F_{2}(E)<\cdots < F_{d}(E)$.

The $\hat{Z}$-box and the associated functions $\{ F_{m}(E)\}$ have the following properties:
\begin{enumerate}
\item[(i)] 
Using Eqs.~(\ref{eq:Q-box}), (\ref{eq:RKK-1}),  (\ref{eq:RKK-2}), and (\ref{eq:z-operator}), we have, for the P-space
eigenstates $\{ |\phi_{k} \rangle \}$ in Eq.~(\ref{eq:p-space-eigen-eq-2}),
\begin{eqnarray}
\label{eq:EKK-KK}
\lefteqn{\sum\limits_{k = 1}^d {\hat Z(E_k )}|{\phi _k } \rangle 
 \langle {\tilde\phi_{k}} |}\nonumber \\
 &=&\sum\limits_{k = 1}^d {\frac{1}{{1 - \hat Q_1 (E_k )}}
 \left[ {R- \hat Q_1 (E_k ) \cdot R} \right]} | {\phi_k }\rangle\langle {\tilde\phi _k } | \nonumber\\
 &=&R. 
 \end{eqnarray}
The above fact means that, replacing $\hat Q(E)$ by $\hat Z(E)$ in 
Eq.~(\ref{eq:RKK-1}), a new solution for the 
effective interaction $R$ can be derived as
\begin{eqnarray}
\label{eq:REKK}
 R_{\tiny\rm EKK}\equiv\sum\limits_{k = 1}^d {\hat Z} (E_k )|{\phi_k}\rangle\langle {\tilde\phi_k}|,
 \end{eqnarray}
and equivalently Eq.~(\ref{eq:p-space-eigen-eq-2}) with the $\hat{Q}$-box is replaced by 
\begin{eqnarray}
 \label{eq:ZE}
 [E_0 P+\hat{Z}(E_k)]|\phi_{k}\rangle=E_{k}|\phi_{k}\rangle.
\end{eqnarray}
From the above relations between two approaches with $\hat{Q}(E)$ and $\hat{Z}(E)$
we may call  $R_{EKK} $ in Eq.~(\ref{eq:REKK}) the extended Krenciglowa-Kuo (EKK) solution. 
In the same way as in Eq.~(\ref{eq:EGK}),
the true eigenvalues $\{ E_{k} \}$ can be given by solving the equations 
\begin{eqnarray}
 \label{eq:FE}
 F_{m}(E)=E, \ \ m=1, 2, \cdots, d.
\end{eqnarray}
\item[(ii)]  Using Eqs.~(\ref{eq:Q-box}), (\ref{eq:Q-E0-derivative}), and (\ref{eq:z-operator}) we can derive a formal expression for the energy derivative of $\hat{Z}(E)$ as
\begin{eqnarray}
\label{eq:Z-diff}
\frac{d\hat{Z}(E)}{dE}&=&\frac{2}{1-\hat{Q}_{1}(E)}\hat{Q}_{2}(E)\nonumber\\
&           &\times[\hat{Z}(E)-(E-E_{0})P].
 \end{eqnarray}
In the paper of Dong {\it et al.}~\cite{DKH2010}, the above expression has also been given for a general case with
the non-degenerate $PH_0P$. 
If $E$ is one of the true eigenvalues $\{E_{k}\}$
 satisfying Eq.~(\ref{eq:ZE}), we see that 
the energy derivative of $\hat Z(E)$ becomes zero at $E=E_{k}$, namely
\begin{eqnarray}
\label{eq:Z-diff-state}
 \left.\frac{d\hat{Z}(E)}{dE}\right|_{E=E_{k}} |\phi_{k}\rangle=0,\ \ k=1, 2, \cdots, d.
 \end{eqnarray}
Resultantly we have for the energy derivative of $F_{m}(E)$
\begin{eqnarray}
\label{eq:Fk-derivative}
\quad \left.\frac{{dF_{m}(E)}}{{dE}} \right|_{E=E_{k}}=0, \ \ k=1, 2, \cdots, d.
\end{eqnarray}
These results for the energy derivatives have been pointed  out in the previous paper \cite{OSKF2010}.
\item[(iii)]  We discuss here some problems associated with the poles of $\hat Q(E)$. 
 First we consider the eigenvalue equation 
for the Q-space Hamiltonian  $QHQ$ written as
\begin{eqnarray}
\label{eq:QHQ-eigenvalue eq}
QHQ|q\rangle =\varepsilon_{q} |q\rangle,
\end{eqnarray}
where $\varepsilon_q$ and $|q \rangle$ are the eigenvalue and the eigenstate, respectively. 
It may be clear from Eq.~(\ref{eq:Q-box}) that  $\hat{Q}(E)$ has a pole at $E=\varepsilon_{q}$.
We define a P-space operator  $\hat{X}_{q}$ with the Q-space eigenstate $|q\rangle$ 
in Eq.~(\ref{eq:QHQ-eigenvalue eq}) as
\begin{eqnarray}
\label{eq:X-operator}
\hat{X}_{q}\equiv PV|q\rangle\langle q|VP.
\end{eqnarray}
We write the  eigenvalue equation for $\hat{X}_{q}$ with an eigenvalue $x_{\mu}$ as
\begin{eqnarray}
\label{eq:X-eigenvalue-eq}
\hat{X}_{q}|{\mu} \rangle&=&x_{\mu}|{\mu} \rangle.
\end{eqnarray}
From the definition of  $\hat{X}_q$ in Eq.~(\ref{eq:X-operator}) we see that $\hat{X}_q$ is Hermitian and
positive semi-definite, that is, $\hat{X}_q$ has positive or zero eigenvalue because of the inequality
$x_{\mu}=|\langle \mu|V|q\rangle|^{2}\ge 0.$
We can further prove that there is only one eigenstate, denoted by  $|\mu_{0} \rangle$,  with a positive eigenvalue
 and that the eigenvalues of all the other eigenstates are zero. The only one positive eigenvalue of Eq.~(\ref{eq:X-eigenvalue-eq})  is given by
\begin{eqnarray}
\label{eq:non-zero-eigenvalue of  X}
x_{\mu_{0}}&=&\sum_{i=1}^{d} |\langle p_{i}|V|q\rangle|^{2},
\end{eqnarray}
where $\{ |p_{i}\rangle; i=1, 2, \cdots, d\}$ are the basis-state vectors of the P space.
The proof is as follows: We condsider  a matrix representation of 
$\hat{X}_{q}$ with the matrix element $(\hat{X}_{q})_{ij}=\langle p_{i}|V|q\rangle\langle q|V|p_{j}\rangle$.
Then we obtain the trace of $\hat{X}_{q}$ as
\begin{eqnarray}
\label{eq:X-matrix-trace}
{\rm Tr}\hat{X}_{q}&=&\sum_{i=1}^{d}|\langle p_{i}|V|q\rangle|^{2}.
\end{eqnarray}
We note that the state vector $|\mu_{0}\rangle$ can be written explicitly as
\begin{eqnarray}
\label{eq:mu-0-state}
|\mu_{0} \rangle&=&\frac{[\langle p_{1}|V|q\rangle,\langle p_{2}|V|q\rangle,\cdots,\langle p_{d}|V|q\rangle ]^{\rm T}}
                              {\sqrt{N}},
\end{eqnarray}
where the symbol {T} means transpose of $(1\times d)$ matrix, and $\sqrt{N}$ is the normalization factor
 with $N\equiv \sum_{i=1}^{d}|\langle p_{i}|V|q\rangle|^{2}$. 
This  state $|\mu_{0} \rangle$ becomes an eigenstate of  $\hat{X}_{q}$ and has a positive eigenvalue as
\begin{eqnarray}
\hat{X}_{q}|\mu_{0} \rangle&=&\left(\sum_{i=1}^{d}|\langle p_{i}|V|q\rangle|^{2}\right)|\mu_{0} \rangle\nonumber\\
\label{eq:mu-0-state-eigen-eq}
&=&x_{\mu_{0}}|\mu_0\rangle.
\end{eqnarray}
The eigenvalue $x_{\mu_{0}}$ coincides with the trace of $\hat{X}_{q}$. 
Recalling a well-known theorem on the trace of a square matrix, we see that the trace of $\hat{X}_q$
should be equal to the sum of the eigenvalues of $\hat{X}_{q}$. It follows immediately that 
all the other eigenvalues except   $x_{\mu_{0}}$ become zero. 
\item[(iv)]  Next, we prove that Eq.~(\ref{eq:ZE}) or (\ref{eq:FE}) has additional solutions other than the 
 true eigenvalues $\{ E_{k}\}$.
We note, in  the vicinity  of a  pole  of  $\hat{Q}(E)$ where $E=\varepsilon_{q}+\Delta$ 
with a small deviation $\Delta$, that $\hat{Q}(E)$ in Eq.~(\ref{eq:Q-box}) is expressed as 
\begin{eqnarray}
\label{eq:Q-operator-near-at-pole}
\hat{Q}(\varepsilon_{q}+\Delta)&=&PVP+\frac{\hat{X}_{q}}{\Delta}+\sum_{q'\ne q}\frac{\hat{X}_{q'}}{\varepsilon_{q}+\Delta-\varepsilon_{q'}}.
\end{eqnarray}
Here we have used the definition of $\hat{X}_{q}$ in Eq.~(\ref{eq:X-operator}). The operations of 
$\hat{Q}(\varepsilon_{q}+\Delta), \hat{Q}_{1}(\varepsilon_{q}+\Delta)$, and  $\hat{Q}_{2}(\varepsilon_{q}+\Delta)$ on the state $|\mu_{0}\rangle$  yield in leading order, respectively
\begin{eqnarray}
\label{eq:Q-mu}
\hat{Q}(\varepsilon_{q}+\Delta)|\mu_{0}\rangle&=&\ \ \frac{x_{\mu_{0}}}{\Delta}|\mu_{0}\rangle,\\
\label{eq:Q1-mu}
\hat{Q}_{1}(\varepsilon_{q}+\Delta)|\mu_{0}\rangle&=&-\frac{x_{\mu_{0}}}{\Delta^{2}}|\mu_{0}\rangle,\\
\label{eq:Q2-mu}
\hat{Q}_{2}(\varepsilon_{q}+\Delta)|\mu_{0}\rangle&=&\ \ \frac{x_{\mu_{0}}}{\Delta^{3}}|\mu_{0}\rangle.
\end{eqnarray}
Using Eq.~(\ref{eq:z-operator}) and the above relations, we obtain at $E=\varepsilon_{q}+\Delta$
\begin{eqnarray}
\label{eq:Z-mu-Delta}
\lefteqn{[ E_{0}+\hat{Z}(\varepsilon_{q}+\Delta) ] |\mu_{0}\rangle }\nonumber \\
&=&\left[
                E_{0}+(\frac{x_{\mu_0}}{\Delta^2})^{-1} \right.
                \nonumber\\
 &&          \hspace{10mm} \times \left.
                           \left[ 
                            \frac{x_{\mu_0}}{\Delta}+\frac{x_{\mu_0}(\varepsilon_{q}+\Delta-E_0 )}{\Delta^2}
                   \right]
     \right] |\mu_{0}\rangle
\nonumber\\
&=&(\varepsilon_{q}+2\Delta ) |\mu_{0}\rangle.\end{eqnarray}
Then we have in the limit of $E \to \varepsilon_{q}$
\begin{eqnarray}
\label{eq:Z-mu}
[E_{0}+\hat{Z}(\varepsilon_{q}) ] |\mu_{0}\rangle&=&\varepsilon_{q}|\mu_{0}\rangle.
\end{eqnarray}
This fact means that the state $|\mu_{0}\rangle$ and the energy $\varepsilon_q$ can be an additional solution to
Eq.~(\ref{eq:ZE}). This also means that
the pole energy $\varepsilon_{q}$ satisfies Eq.~(\ref{eq:FE}). 
\item[(v)] 
We further consider the energy derivatives of $\hat{Z}(E)$ and $F_{m}(E)$ at the pole positions.
Subtraction of Eq.~(\ref{eq:Z-mu}) from Eq.~(\ref{eq:Z-mu-Delta}) follows
\begin{eqnarray}
\label{eq:Z-mu-derivative}
\left.\frac{d\hat{Z}(E)}{dE}\right|_{E=\varepsilon_{q}} |\mu_{0}\rangle&=&2|\mu_{0}\rangle.
\end{eqnarray}
We can derive the above result, in another way,  by using Eqs.~(\ref{eq:Z-diff}) and (\ref{eq:Z-mu})
\begin{eqnarray}
\lefteqn{\left.\frac{d\hat{Z}(E)}
              {dE}\right|_{E=\varepsilon_{q}+\Delta} |\mu_{0}\rangle }\nonumber \\
&=&\frac{2}{1-\hat{Q}_{1}(\varepsilon_q +\Delta)}\hat{Q}_2 (\varepsilon_q +\Delta)\cdot \Delta |\mu_{0}\rangle .
\end{eqnarray}
Then, using Eqs.~(\ref{eq:Q1-mu}) and (\ref{eq:Q2-mu}) and  taking the limit of $\Delta \to 0$, we obtain
\begin{eqnarray}
\lefteqn{\left.\frac{d\hat{Z}(E)}
              {dE}\right|_{E=\varepsilon_{q}} |\mu_{0}\rangle }\nonumber \\
                 &=&\lim_{\Delta \to 0}  
                       2\cdot \left( \frac{x_{\mu_0}}{\Delta^2} \right)^{-1}
                    \cdot \frac{x_{\mu_0}}
                                    {\Delta^3}
                    \cdot \Delta |\mu_{0}\rangle\nonumber\\
         &=&2|\mu_{0}\rangle.
\end{eqnarray}
As a consequence we have
\begin{eqnarray}
\label{eq:F-derivative-at-pole}
 \left.\frac{dF_{m}(E)}{dE} \right|_{E= \varepsilon_{q}}&=&2.
 \end{eqnarray}
This property of the energy derivative of $F_{m}(E)$ at a pole position is quite contrast to that for 
a true eigenvalue in Eq.~(\ref{eq:Fk-derivative}).
These properties  of the energy derivative of  $F_{m}(E)$ given in Eqs.~(\ref{eq:Fk-derivative}) and (\ref{eq:F-derivative-at-pole}) 
can be used to distinguish the pole energy solutions from those of the
true eigenvalues of $H$.  It is also noted, from  Eqs.~(\ref{eq:Z-mu}) and (\ref{eq:F-derivative-at-pole}),
that the function
$F_{m}(E)$ is a well-behaved function of $E$ and has no singularity even at the pole energies.
\end{enumerate}
\subsection{\label{iteration}Iteration methods for the effective interaction} 
Iteration methods have been employed quite often for obtaining the effective interaction or equivalently for 
determining the true eigenvalues of $H$. In the previous subsections we have  given four basic equations,
namely Eqs.~(\ref{eq:p-space-eigen-eq-2}) and (\ref{eq:EGK}) for the KK and Eqs.~(\ref{eq:ZE}) and (\ref{eq:FE}) for the EKK methods.
In the KK approach Eqs.~(\ref{eq:p-space-eigen-eq-2}) and (\ref{eq:EGK}) lead, respectively, to the following iterative equations:
\begin{eqnarray}
\label{eq:p-space-eigen-eq-2-iteration}
[E_{0}P+{\hat Q} (E_k^{(n)} )]|{\phi_k^{(n+1)}}\rangle&=&E_{k}^{(n+1)}|{\phi_k^{(n+1)}}\rangle
 \end{eqnarray}
and 
\begin{eqnarray}
\label{eq:EGK-iteration}
E^{(n+1)}&=&G_{m}(E^{(n)}), \ \ m=1, 2, \cdots, d.
\end{eqnarray}
In the same way two iterative equations can be derived from Eqs.~(\ref{eq:ZE}) and (\ref{eq:FE}) for the EKK method, respectively as 
\begin{eqnarray}
\label{eq:Z-p-space-eigen-eq-2-iteration}
[E_{0}P+{\hat Z} (E_k^{(n)} )]|{\phi_k^{(n+1)}}\rangle&=&E_{k}^{(n+1)}|{\phi_k^{(n+1)}}\rangle
 \end{eqnarray}
and 
\begin{eqnarray}
\label{eq:EFK-iteration}
E^{(n+1)}&=&F_{m}(E^{(n)}), \ \ m=1, 2, \cdots, d.
\end{eqnarray}
The iterative equation (\ref{eq:p-space-eigen-eq-2-iteration}) has long been used as a standard method in the KK approach.
The convergence condition has been investigated in many of the theoretical and numerical studies~\cite{KK74,Suzuki-Lee80,GLS94}.
 From these studies it has been
known that, if the iteration converges, the KK approach in Eq.~(\ref{eq:p-space-eigen-eq-2-iteration}) 
derives $d$ eigenvalues of the eigenstates of $H$ with the largest P-space overlaps. 
However, it has also been known that the iteration in 
Eq.~(\ref{eq:p-space-eigen-eq-2-iteration}) does not always converge. 
Recently  Takayanagi~\cite{Takayanagi2010} has pointed out an exceptional case
that the KK method does not reproduce states which have the largest P-space overlaps
even if the iteration converges.
Although the KK iteration method has been applied widely and has brought remarkable results in actual 
calculations, the rigorous condition of convergence for 
Eq.~(\ref{eq:p-space-eigen-eq-2-iteration}) has not yet been made clear.

The convergence condition for the iteration in Eq.~(\ref{eq:EGK-iteration}) requires mathematically  that 
the energy derivative of $G_m(E)$ at the solution $ E=E_k$ should satisfy $|dG_m(E)/dE|_{E=E_k}<1$.
In general, this convergence condition  is considered to be satisfied by certain of the true eigenvalues of $H$. It implies that this iteration method is restrictive for the purpose of reproducing the solutions of $\{ E_{k}\}$ as many as possible.

The iteration in Eq.~(\ref{eq:Z-p-space-eigen-eq-2-iteration}) leads to a new scheme of the calculations
of $\{ E_k \}$ and $R_{\rm EKK}$ in Eq.~(\ref{eq:REKK}). As in the case of the KK iteration in
 Eq.~(\ref{eq:p-space-eigen-eq-2-iteration}), the convergence condition in Eq.~(\ref{eq:Z-p-space-eigen-eq-2-iteration}) is quite complicated and has not yet been
 made clear. However, the iteration in Eq.~(\ref{eq:Z-p-space-eigen-eq-2-iteration}) would be
 applicable in  actual calculations. In the paper of Dong {\it et al.}~\cite{DKH2010}, they have employed this iteration scheme to a 
 model problem and actual calculations of the shell-model effective interaction. They have concluded that
 the two iterations in Eq.~(\ref{eq:p-space-eigen-eq-2-iteration})  for the KK and 
Eq.~(\ref{eq:Z-p-space-eigen-eq-2-iteration}) for the EKK methods  are both suitable and efficient.

\begin{figure}[t]
\includegraphics[width=.360\textheight]{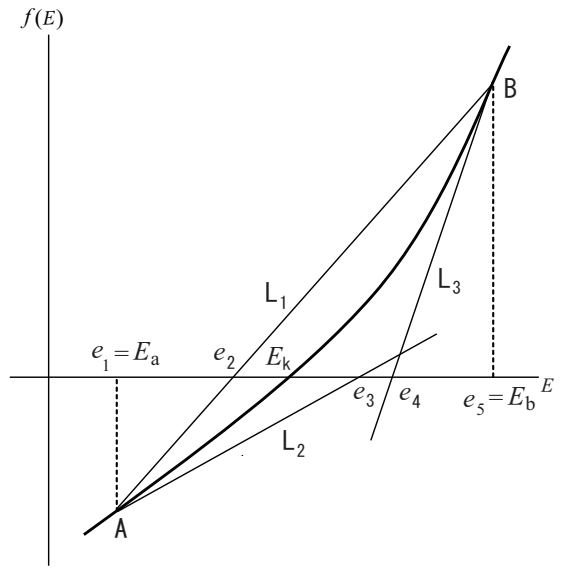}
\caption{\label{fig-1new} Graphical representation of the modified secant method. 
The ${\rm L}_1$ is the straight line passing through the points A and B.
The ${\rm L}_2$ and ${\rm L}_3$ are the tangents to the function $f(E)$
 at the points A and B, respectively. See text for detailed explanation.}
\end{figure}


Another iterative equations in the EKK method is given in Eq.~(\ref{eq:EFK-iteration}).
The convergence condition for this iteration is given by $|dF_m(E)/dE|_{E=E_k}<1$.
It is clear from the property of the energy derivative of $F_m(E)$ in Eq.~(\ref{eq:Fk-derivative}) that 
the convergence condition is satisfied for any of the true eigenvalues $\{ E_k \}$. 
The iteration in Eq.~(\ref{eq:EFK-iteration}) converges surely to one of $\{ E_k \}$.
We also note that this iteration in Eq.~(\ref{eq:EFK-iteration}) never reaches the pole-energy solutions due to the property in 
Eq.~(\ref{eq:F-derivative-at-pole}) of the energy derivative of $F_m(E)$.
We can show that the iteration in Eq.~(\ref{eq:EFK-iteration}) is equivalent to that
in the Newton-Raphson method which is often used to solve non-linear equations.
 The Newton-Raphson method has been known to derive quadratic convergence, that is, 
the number of correct digits is doubled at each step of iteration. On the contrary  the usual methods,
 including the KK and the LS methods, derive single (or linear) convergence. 
The iteration in  Eq.~(\ref{eq:EFK-iteration}) guarantees surely faster convergence than the usual iterations 
 given so far.
\subsection{\label{graphical}Graphical method for effective interaction with the $\hat{Z}$-box}
We present a new method for solving Eq.~(\ref{eq:FE}) derived on the basis of the $\hat{Z}$-box and its associated 
functions $\{ F_m(E)\}$. The solutions  can be obtained by finding the energies of the intersections
of two graphs, $y=F_{m}(E)$ and $y=E$. As we have already shown, the $\hat{Z}$-box has no singularities at the pole positions of the $\hat{Q}$-box.
Thus, $\{ F_{m}(E) \}$ can be considered to be well-behaved functions of $E$.
These characteristics of the functions  $\{ F_{m}(E) \}$ enable us to employ some of the mathematically well-established 
methods for solving non-linear equations.

The procedure of solving Eq.~(\ref{eq:FE}) in the present approach, which we call  the graphical method, 
is given as follows:
We first draw graphs of $y=F_{m}(E)$ for $m=1, 2, \cdots, d$ and $y=E$. 
The energies at the intersections of these  graphs 
become the solutions to $F_{m}(E)=E$. Furthermore, the energy derivatives of $\{ F_{m}(E) \}$ at the intersections distinguish
the pole-energy solutions $\{ \varepsilon_q\}$ from those of the true eigenvalues $\{ E_{k}\}$. As was proved in the 
previous subsections, the energy derivatives, denoted by $\{ F'_{m}(E)\}$, should be  $F'_{m}(E)=0$ for $E=E_k$ and  $F'_{m}(E)=2$ for $E=\varepsilon_q$. With these simple regulations for $\{ F'_{m}(E)\}$ we can easily specify the solutions of  the true eigenvalues
$\{ E_k\}$.

We next figure out, from the intersections of the graphs, roughly estimated  solutions for $\{ F_{m}(E)=E\}$. 
Starting with these approximate
solutions, we proceed to make further precise calculations for the solutions. For this purpose we employ a combined
method of the secant and the Newton-Raphson methods~\cite{secant-method}. The algorithm of this method is as follows: Let us define  a function $f(E)$ as 
$f(E)\equiv E-F(E)$, where $F(E)$ is one of the functions $\{ F_{m}(E) \}$. The solutions to $F_{m}(E)=E$ are obtained as the roots of $f(E)=0$. 
Suppose that $E_a$ and $E_b$ be two values which bracket one of the roots and satisfy $E_a<E_b$ and $f(E_a)f(E_b)<0$. We suppose also that $f(E)$ be a monotone
function on the interval $E_a <E<E_b$. These suppositions on $E_a$ and $E_b$ follow that only one root exists on
the interval $E_a <E<E_b$. As shown in Fig.~\ref{fig-1new}, we further determine five points according to 
\begin{eqnarray}
\label{eq:e1}
e_1&=& E_a,\\
\label{eq:e2}
e_2&=&\frac{E_a \cdot f(E_b)-E_b \cdot f(E_a)}{f(E_b)- f(E_a)},\\
\label{eq:e3}
e_3&=&E_a-\frac{f(E_a)}{f'(E_a)},\\
\label{eq:e4}
e_4&=&E_b-\frac{f(E_b)}{f'(E_b)},\\
\label{eq:e5}
e_5&=& E_b.
\end{eqnarray} 
The value $e_2$ in Eq.~(\ref{eq:e2}) is the better approximate solution in the usual secant method. The $e_3$ and $e_4$ are the approximate
values in the Newton-Raphson method. We can easily select two values, 
$e_i$ and $e_j$,  among $\{ e_1, e_2, \cdots,e_5\}$ such that they are neighboring on the $E$-axis and satisfy $E_a \le e_i <e_j \le E_b$ and $f(e_i) f(e_j)<0$.
\begin{table}[t]
\caption{\label{table-1} Convergence of the eigenvalues of the lowest-lying and second lowest-lying states obtained 
in the iterations $E^{(n+1)}=G_{m}(E^{(n)})$  and $E^{(n+1)}=F_{m}(E^{(n)})$ for  the KK and for  the EKK methods,
 respectively, for  the model Hamiltonian with the strength $x=0.05$. Correct digits in  the KK and the EKK methods are given for $n$,  namely the number of iterations. The starting energies 
 $(E_{1}^{(1)}, E_{2}^{(1)})$ are taken to be $(0.0, 0.0)$. The notation c indicates convergence to more than fifteen decimal places. The exact eigenvalues here are $E_{1}=0.8904504858869942$ and $E_{2}=2.2156808150096040$.}
\begin{ruledtabular}
    \begin{tabular}{clll}
$n$ &correct digits (KK)&correct digits ({\rm EKK}) \\ \hline
1  & 0.9$\cdots$            & 0.89$\cdots$    \\
2  & 0.88$\cdots$          & 0.890450$\cdots$   \\
3  & 0.8904$\cdots$         & 0.89045048588699$\cdots$\\
4  & 0.890450$\cdots$       &c\\
5  & 0.8904504$\cdots$     &c\\
6  & 0.890450485$\cdots$    &c\\
7  & 0.8904504858$\cdots$  &c   \\ \hline
1  & 2.2$\cdots$            & 2.2$\cdots$    \\
2  & 2.21$\cdots$          & 2.21568$\cdots$   \\
3  & 2.215$\cdots$         & 2.2156808150$\cdots$\\
4  & 2.2156$\cdots$       &c              \\
5  & 2.21568$\cdots$     & c             \\
6  & 2.215680$\cdots$    & c \\
7  & 2.2156808$\cdots$  & c \\
8  & 2.21568081$\cdots$  & c \\
9  & 2.215680815$\cdots$  & c  \\
    \end{tabular}
\end{ruledtabular}
\end{table}
These conditions mean that one of the roots exists on the interval $e_i <E<e_j$. We then replace  $E_a$ and $E_b$ as $E_a=e_i $ and $E_b=e_j $ and repeat the procedure again. 
This modified secant method derives surely one of the roots on the interval $E_a<E<E_b$. We also note  
that this method will lead to fast convergence, because we calculate approximate solutions according to  the Newton-Raphson method.

We here wish to emphasize that this method is not an iteration method. The procedure in this method 
always derives  a convergent solution to $f(E)=0$ on the interval $E_a<E<E_b$ and never reaches any of 
the solutions outside the
interval. Therefore, we can always control convergence and select solutions to be reproduced
 by selecting starting values $E_a$ and $E_b$ properly. Only one problem in the present method is how to choose $E_a$ and $E_b$.
Then the use of the graphs of $\{ y=F_m(E)\}$ would be helpful. If the graphs are drawn
 accurately, we can determine easily these starting energies $E_a$ and $E_b$.

\begin{figure}[t]
\includegraphics[width=.360\textheight]{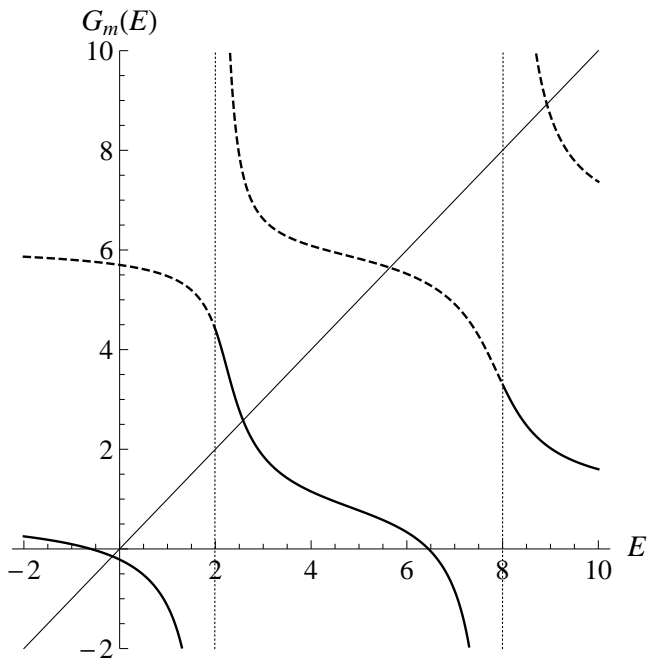}
\caption{\label{fig-2} Dependence of $\{G_{m}(E)\}$ on $E$ with $x=0.2$.
The graphs of $ y=G_{1}(E)$ and $y=G_{2}(E)$ are shown in solid and broken lines, respectively. The graph of $y=E$ is also shown.}
\end{figure}

\begin{figure}[t]
\includegraphics[width=.360\textheight]{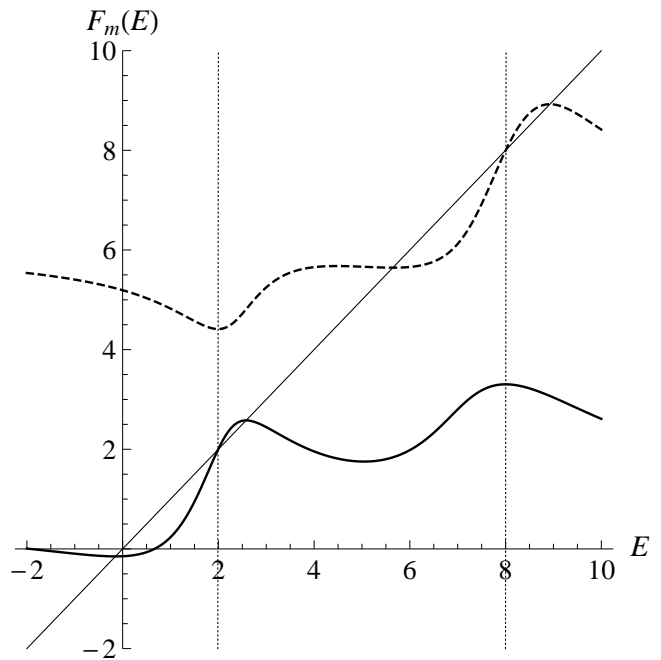}
\caption{\label{fig-3} Dependence of $\{ F_{m}(E)\}$ on $E$ with $x=0.2$.
The graphs of $ y=F_{1}(E)$ and $y=F_{2}(E)$ are shown in solid and broken lines, 
respectively. The graph of $y=E$ is also shown.}
\end{figure}

\section{\label{test}Test calculations}
In order to assess the present method we consider a model
problem for which exact results can be obtained easily. The model we adopt here~\cite{GLS94} is a slightly modified version of the one which was introduced many years ago to study the intruder state problem~\cite{SW72,Hoffmann73}.  
The dimensions of the entire space and the P space are four and two ($d=2$), respectively.
The degenerate unperturbed energy is taken to be $E_0=1$, and the interaction $V$ is given with the coupling strength $x$. 
The relevant matrix elements are given by
\begin{eqnarray}
\label{eq:model-hamiltonian-1}
PH_{0}P 
  &=&\left( 
      {\begin{array}{*{20}c}
       1 & 0  \\
       0 & 1  \\
       \end{array}} 
     \right),\ \nonumber \\
PVP 
  &=& \left( 
    {\begin{array}{*{20}c}
       0  & {5x}  \\
     {5x} & {25x}  \\
     \end{array}} \right),\;
PVQ 
  = \left( 
     {\begin{array}{*{20}c}
      { - 5x} & {5x}     \\
      {5x}    & { - 8x}  \\
     \end{array}} 
    \right),
\end{eqnarray}
and
\begin{eqnarray}
\label{eq:model-hamiltonian-2}
 QHQ 
  = \left( 
     {\begin{array}{*{20}c}
       {3 - 5x} & x        \\
          x     & {9 - 5x} \\
      \end{array}} 
    \right). 
\end{eqnarray}

We present in Table~\ref{table-1} the results of the iterative calculations, based on Eq.~(\ref{eq:EGK-iteration}) for the KK
 and Eq.~(\ref{eq:EFK-iteration})
for the EKK methods. The coupling strength $x$ in the model Hamiltonian is taken to be $x=0.05$. Table~\ref{table-1}
shows 
the results for the lowest and second-lowest eigenvalues of $H$. We observe that the convergence  in the EKK method is much faster than that in the KK method. The EKK iteration reaches convergence to fourteen decimal places after three iterations. It is also impressive that the KK iteration shows linear and steady convergence. 
As has already been discussed
in Sec.~II, the iteration in the EKK method is essentially equivalent to that 
in the Newton-Raphson method which derives quadratic convergence. The calculation in the EKK method verifies this
theoretical prediction that the number of correct digits is doubled at each iteration. 


\begin{table}[t]
\caption{\label{table-2} Convergence of the solutions  obtained with the graphical method for the model Hamiltonian with the strength $x=0.05.$ 
The $E_i$ and $F'(E_i)$ for $1\le i \le 4$ are the true eigenvalues of $H$ and the energy derivatives of 
$\{F_m (E)\}$ at $E=E_i$, respectively.  The $\varepsilon_{q_{1}}$ and $\varepsilon_{q_{2}}$are the solutions for the pole energies.
The $F'_m(\varepsilon_{q_{1}})$ and  $F'_m(\varepsilon_{q_{2}})$ are the energy derivatives of $\{F_m (E)\}$ at $E=\varepsilon_{q_{1}}$ and $\varepsilon_{q_{2}}$, respectively. The notation c indicates convergence to more than six decimal places.
The number of repeats in the modified secant method is given by $n$. The exact values for the solutions
are also given.}
\begin{ruledtabular}
    \begin{tabular}{clclclc|c}
$n$                 & $E_{1}$       & $F'(E_{1})$   & $E_{2}$& $F'(E_{2})$ \\
\hline
1                    & 0.890462    & $<10^{-6}$   & 2.215995  & 0.00011\\
2                    & 0.890450    & $<10^{-6}$   & 2.215681  &  $<10^{-6}$ \\
3                    & c            & $<10^{-6}$     & c            &  $<10^{-6}$ \\
Exact              & 0.890450    & 0.0             &  2.215681  &    0.0       \\
$(E_{a}, E_{b})$   & (0.8, 0.9)    &                  &  (2.2, 2.3)  &            \\
\hline
$n$                 & $E_{3}$         & $F'(E_{3})$ &  $E_{4}$ & $F'(E_{4})$ \\
\hline
1                    & 2.837465       & 0.41335    & 8.781060 & 0.04125 \\
2                    & 2.860730       & 0.02242    & 8.781706 & 0.00014\\
3                    & 2.862158       & 0.00004    & 8.781708 & $<10^{-6}$\\
4                    & 2.862161       & $<10^{-6}$ & c          & $<10^{-6}$\\
Exact              & 2.862161       & 0.0           & 8.781708 &    0.0     \\
$(E_{a}, E_{b})$   & (2.8, 2.9)       &                &  (8.78, 8.80) &           \\
\hline
$n$          & $\varepsilon_{q_1}$ & $F'(\varepsilon_{q_1})$ & $\varepsilon_{q_2}$ & $F'(\varepsilon_{q_2})$ \\
\hline
1             & 2.750350             & 1.98829                    & 8.751088             &  1.95724                   \\
2             & 2.749587             & 1.99995                     & 8.750417            & 1.99999                    \\
3             & 2.749583             & 2.00000                     & c                      & 2.00000                     \\
Exact       & 2.749583             & 2.0                           & 8.750417             & 2.0                     \\
$(E_{a}, E_{b})$ & (2.70, 2.75)     &                                & (8.70, 8.77)           &                             \\
    \end{tabular}
\end{ruledtabular}
\end{table}


We depict, in Figs.~\ref{fig-2} and \ref{fig-3}, the dependence of the functions $\{ G_m(E); m=1, 2\}$ and 
$\{ F_m(E); m=1, 2\}$, respectively, on the energy variable $E$ with the coupling strength $x=0.2$.
One observes in Fig.~\ref{fig-2} that there are two poles in the graphs of $\{G_m(E)\}$ associated with the $\hat{Q}$-box.
On the other hand, the poles disappear in the graphs of $\{F_m(E)\}$ associated with the $\hat{Z}$-box.
One sees four intersections of $y=G_m(E)$ and $y=E$,
 and six intersections of $y=F_m(E)$ and $y=E$ for $m=1, 2$.
Among the six intersections two of these correspond to the pole energies, the eigenvalues of $QHQ$.
The energy derivatives $\{F'_m(E)\}$ at the intersections should be zero for the true eigenvalues of $H$, which are shown in Eq.~(\ref{eq:Fk-derivative}). On the contrary the energy derivatives $\{ F'_m(E)\}$ at the intersections should be two
 for the solutions of the pole energies.
The number of the intersections in the graphs of $\{ G_m(E)\}$ and $\{ F_m(E)\}$ is verified as already predicted
in Sec.~II. The theoretical predictions on the energy derivatives $\{F'_m(E)\}$ of these functions are also verified as shown in Fig.~3.

We can obtain, from these graphs in Figs.~2 and 3, much information on the convergence conditions in some
iteration methods. One observes in Fig.~\ref{fig-2} that the energy derivatives of the first and  third low-lying 
intersections are less than one and those of the second and  fourth are larger than one. From these observations
we may say that the iteration in Eq.~(\ref{eq:EGK-iteration}) cannot reproduce 
the second and fourth eigenvalues of $H$.
On the other hand, one sees from Fig.~\ref{fig-3} that there are  four intersections with zero energy derivatives. This means that the iteration in Eq.~(\ref{eq:EFK-iteration}) in the EKK method always converges to any of the four true eigenvalues of $H$.

We made another numerical calculation by using the $\hat{Z}$-box and the associated function $F_{m}(E)$ 
in the modified secant method.
Tables~\ref{table-2}, III, and IV show the results for the coupling strength $x=0.05, 0.1$, and $0.2$, respectively. 
The starting energies $E_a$ and $E_b$ are taken as 
approximate solutions for the intersections of the graphs of $\{ y=F_{m}(E)\}$ and $y=E$. The results show that all the solutions to 
Eq.~(\ref{eq:FE}) are reproduced. They include the four true eigenvalues $\{ E_1,\cdots, E_4\}$ of $H$ and the two pole energies $\{ \varepsilon_{q_1},\varepsilon_{q_2}\}$. The values of the energy derivatives $\{ F'_{m}(E)\}$ at the solutions were also calculated.
 As already proved in Sec.~II theoretically, the energy derivative at the true eigenvalues of $H$ should be zero and,
 on the other hand, two at 
the pole energies. These theoretical predictions  are also confirmed numerically. From this difference of the derivatives of
$F_{m}(E)$ we can easily classify the solutions into two parts, the true eigenvalues of $H$ and the pole energies.


\begin{table}[t]
\caption{\label{table-3} Convergence of the solutions  obtained with the graphical method for the model Hamiltonian with the strength $x=0.1.$
As for the notations see Table II.}
\begin{ruledtabular}
    \begin{tabular}{clclclc|c}
$n$                 & $E_{1}$        & $F'(E_{1})$ & $E_{2}$ & $F'(E_{2})$ \\
\hline
1                    & 0.648389      & 0.00002    & 2.552215 & 0.04647 \\
2                    & 0.648250     & $<10^{-6}$ & 2.553840 & 0.00016\\
3                    & c             & $<10^{-6}$ &2.553845  & $<10^{-6}$\\
Exact              & 0.648250      & 0.0          & 2.553845 & 0.0     \\
$(E_{a}, E_{b})$   & (0.6, 0.7)       &              & (2.55, 2.65) &         \\
\hline
$n$                 & $E_{3}$       & $F'(E_{3})$ & $E_{4}$  & $F'(E_{4})$ \\
\hline
1                    & 3.649852    & 0.00005    & 8.647500 & 0.00392 \\
2                    &3.650111     & $<10^{-6}$& 8.647794 & $<10^{-6}$ \\
3                    &c                & $<10^{-6}$ &         & $<10^{-6}$ \\
Exact              &3.650111    &     0.0        & 8.647794 &    0.0     \\
$(E_{a}, E_{b})$   & (3.6, 3.7)    &                & (8.63, 8.65) &           \\
\hline
$n$               & $\varepsilon_{q_1}$ &$F'(\varepsilon_{q_1})$ & $\varepsilon_{q_2}$ & $F'(\varepsilon_{q_2})$ \\
\hline
1                  & 2.498646             & 1.99121             &  8.501897               & 1.99693 \\
2                  & 2.498334            & 1.99999             &  8.501666                 & 2.00000 \\
3                  & c                      & 2.00000              & c                           & 2.00000 \\
Exact            & 2.498334              &  2.0                  &  8.501666                &  2.0     \\
$(E_{a}, E_{b})$& (2.48, 2.50)            &                         &   (8.50, 8.52)           &           \\
    \end{tabular}
\end{ruledtabular}
\end{table}


\begin{table}[t]
\caption{\label{table-4} Convergence of the solutions  obtained with the  strength $x=0.2.$ 
As for the notations see Table II.}
\begin{ruledtabular}
    \begin{tabular}{clclclc|c}
$n$                 & $E_{1}$      & $F'(E_{1})$ & $E_{2}$    & $F'(E_{2})$ \\
\hline
1                    & -0.149272   & 0.00008   & 2.577187  & 0.00535 \\
2                    & -0.149586  & $<10^{-6}$ & 2.579424  & $<10^{-5}$ \\
3                    & c              & $<10^{-6}$ & 2.579425   & $<10^{-6}$\\
Exact              & -0.149586  & 0.0           & 2.579425   & 0.0         \\
$(E_{a}, E_{b})$   & (-0.2, -0.1)  &               & (2.5, 2.6)  &            \\
\hline
$n$                 & $E_{3}$       & $F'(E_{3})$ & $E_{4}$   & $F'(E_{4})$ \\
\hline
1                    & 5.645253    & 0.00003     & 8.923645 &  0.00244 \\
2                    & 5.645051    & $<10^{-6}$ & 8.925109 & $<10^{-5}$\\
3                    & c         & $<10^{-6}$  & 8.925110 &$<10^{-6}$\\
Exact              & 5.645051   &     0.0         & 8.925110 &    0.0     \\
$(E_{a}, E_{b})$   & (5.6, 5.7)    &                 & (8.9, 9.0) &           \\
\hline
$n$              & $\varepsilon_{q_1}$ &  $F'(\varepsilon_{q_1})$ &$\varepsilon_{q_2}$ & $F'(\varepsilon_{q_2})$ \\
\hline
1                  & 1.993869                   & 1.99886                &   8.007196         & 1.99921 \\
2                  & 1.993341                 & 2.00000                  &   8.006659         & 2.00000 \\
3                  & c                           & 2.00000                   &  c                    & 2.00000 \\
Exact            & 1.993341                  &  2.0                          &   8.006659      &  2.0     \\
$(E_{a}, E_{b})$& (1.9, 2.0)                    &                               &   (8.0, 8.1)        &           \\
    \end{tabular}
\end{ruledtabular}
\end{table}


As a whole the convergence rates are reasonable in the cases with the coupling strength $x=0.05, 0.1,$ and $0.2$. Three steps of the calculations in the 
modified secant method are enough for yielding the results with accuracy to six decimal places.
However, the convergence depends strongly on the choice of the starting energies $(E_a, E_b)$. When the spacing of two
solutions among  $\{ E_k\}$ and $\{ \varepsilon_q\}$ is very narrow, we need to draw graphs accurately enough
for finding approximate values of $\{ E_a, E_b\}$ to bracket each of the solutions.

\section{\label{concluding}Concluding remarks}
We have introduced a new vertex function called the $\hat{Z}$-box, which is an operaor defined in a model space and a function of an energy $E$. The $\hat{Z}$-box is
constructed in terms of the $\hat{Q}$-box and its energy derivative originated by Kuo {\it et al}. We have proved that a new expression of the effective interaction can be derived  by replacing the $\hat{Q}$-box
by the $\hat{Z}$-box in the KK method. With the $\hat{Z}$-box we have also introduced a set of scalar functions
 $\{ F_m(E)\}$.
It has been shown that the $\hat{Z}$-box and the associated functions $\{F_m(E)\}$
 have the following properties:
\begin{enumerate}
\item[(i)]  The true eigenvalues $\{ E_k \}$ of the original Hamiltonian $H$ can be given by the roots of a set 
of equations $\{ F_m(E)=E\}$. These equations  have the roots, as additional solutions, at the pole energies 
$\{ \varepsilon_q\}$ of the $\hat{Q}$-box.
\item[(ii)]  The $\hat{Z}$-box and the functions $\{ F_m(E)\}$ have no singularities even at the poles of the $\hat{Q}$-box,
and they are well-behaved functions of $E$.
\item[(iii)]  The derivatives of $\{F_m(E)\}$ at the solutions for $E$ take two values, {\it i.e.}, zero at the true eigenvalues 
$\{ E_k\}$ of $H$, and two at the pole energies $\{ \varepsilon_q\}$.
\end{enumerate}
On the basis of the properties (i), (ii), and (iii), we have proposed a new iteration scheme written as
$E^{(n+1)}=F_{m}(E^{(n)})$. This iteration always converges and reproduces the 
true eigenvalues $\{ E_k\}$ of $H$. Since the energy derivatives of  $\{ F_m(E)\}$
at the solutions for $\{ E_k\}$ are always zero, this method can be understood to be equivalent to
the Newton-Raphson method used to solve non-linear equations. The Newton-Raphson iteration derives quadratic
convergence. Therefore, we can expect that this new iteration leads to fast convergence. We have 
carried out a test calculation and confirmed the quadratic convergence.

As another method for solving a set of equations
$\{ F_m(E)=E\}$, we have proposed a new non-iterative method which we call the graphical method.
The solutions of these equations can be obtained as the energies at the intersections
of the graphs $\{ y=F_m(E)\}$ and $y=E$. Using the property (iii) we can classify the intersections into
 two parts, {\it i.e.}, those for the solutions of the true eigenvalues $\{ E_k\}$ and of the pole energies
 $\{ \varepsilon_q\}$. These graphs make us possible to estimate roughly the positions of the roots.
 With approximate solutions obtained from the graphs we proceed to make more accurate 
 calculations of the solutions, where we have employed a modified secant method
 which is a combined method of the secant  and the Newton-Raphson methods. This method has been shown 
 to provide a suitable scheme for obtaining accurate results if we start with approximate energies close to the solutions.
 
 We have made test calculations to assess the graphical method.
 We have confirmed numerically that the modified secant method reproduces successfully  all the solutions, including the true eigenvalues $\{ E_k\}$ of $H$ and the pole energies $\{ \varepsilon_q\}$. 
 The theoretical predictions in the property (iii) on the energy derivatives of $\{ F_m(E)\}$ are also verified numerically.
 We wish to note that the graphical method, implemented by the modified secant method, yields always convergent results, where we do not need any information on the eigenstates and/or the eigenvalues of $H$ such as P-space overlaps and/or energy spacings.
 
 We may conclude that the present approach, the graphical method with the vertex function $\hat{Z}(E)$, would be promising in resolving some of the difficulties in the derivation of the effective interaction.

\begin{acknowledgments}

We are grateful to T.~T.~S.~Kuo, D.~Rowe, and L.~Coraggio for their  helpful and stimulative discussions.
We also thank H.~Kamada, H.~Kimura, and S.~Maeda for their computational advices.

\end{acknowledgments}

\end{document}